\documentclass[twocolumn,prb,showpacs,floatfix,amsfonts]{revtex4}
\pdfoutput=1
\usepackage{textcomp}
\usepackage{amssymb}
\usepackage{graphicx}

\usepackage[pdftex]{hyperref}

\begin{document}

\title{Unusual reflection of electromagnetic radiation
from a stack of graphene layers at oblique incidence }

\author{Yu. V. Bludov, N. M. R. Peres, M. I. Vasilevskiy }

\affiliation{Department of Physics and Center of Physics, University of Minho,
P-4710-057, Braga, Portugal}

\email{bludov@fisica.uminho.pt}

\begin{abstract}
We study the interaction of electromagnetic (EM) radiation with single-layer graphene and a stack of parallel graphene sheets at arbitrary
angles of incidence. It is found that the behavior is qualitatively different for transverse
magnetic (or $p-$polarized) and transverse electric (or $s-$polarized) waves. In particular, the absorbance of single-layer graphene attains minimum (maximum) for $p$ ($s$) polarization, at the angle of total internal reflection when the light comes from a medium with a higher dielectric constant. In the case of equal dielectric constants of the media above and beneath graphene, for grazing incidence graphene is almost 100\% transparent to $p-$polarized waves and acts as a tunable mirror for the $s-$polarization. These effects are enhanced for the stack of graphene sheets, so the system can work as a broad band polarizer.
It is shown further that a periodic stack of graphene layers has the properties of an one-dimensional photonic crystal, with gaps (or stop--bands) at certain frequencies.
When an incident EM wave is reflected from this photonic crystal, the tunability of the graphene conductivity renders the possibility of controlling the gaps, and the structure can operate as a tunable spectral--selective mirror.
\end{abstract}


\maketitle

\section{Introduction}

Electromagnetic (EM) metamaterial engineering yields specific optical properties which do not exist in natural
materials\cite{Engheta2006}.
  These properties include EM energy concentration in sub-wavelength regions and radiation guiding\cite{Bozhevolnyi2013}, enhanced absorption\cite{Kravets2008}, reflection\cite{Joannopoulos2008} and transmission\cite{GarciadeAbajo2007}, colour filtering\cite{TingXu2010}, {\it etc.}
An important and prominent example of
metamaterials and their specific properties are photonic crystals
(PCs)\footnote {Although photonic crystals are different from the "classical" metamaterials (devices with negative refractive index) since the wavelength of operation is comparable to their structural period, the term "metamaterial" nowadays is applied to any artificial structure with designed material (in our case, optical) properties.},
  where the propagation of electromagnetic waves of certain frequencies,
belonging to gaps (or stop-bands) in the spectrum, can be prohibited, or
allowed in certain directions only\cite{Joannopoulos2008}. Thus,
the so called three-cylinder structure\cite{Yablonovitch1991} was the
first experimental realization of full photonic band gap, where the propagation
of electromagnetic waves is not possible in any direction. The photonic
band-gap structure of PC resembles and appears in full analogy with
the electronic one in solid-state crystals.

In the metamaterial\textasciiacute{}s engineering it is useful to implement some
tools for adjusting their EM properties, thus achieving the
\textit{tunability}. Tunable metamaterials allow for continuous variation
of their properties through a certain external influence (for review
see, e.g. Refs.\cite{Shalaev2011,Vendik2012,Zheludev2012}). Among the
possible instruments to achieve the PC\textasciiacute{}s dynamical
tunability we can mention the optical beam intensity in a nonlinear material
\cite{Chen2011}, electric field in ferroelectrics\cite{Figotin1991}, applied voltage in
liquid crystals\cite{ozaki2002,Li2007}, magnetic field in ferromagnets or ferrimagnets
\cite{Figotin1991,PRB2000}, and mechanical force for changing the PC period\cite{APL2004}.
There are also possible tuning mechanisms in crystalline colloidal arrays of high refractive index particles
\cite{advmat2012}, magnetic fluids\cite{magFluid2012} or superconductors \cite{savelev2005,savelev2006}.

The two-dimensional carbon material graphene possesses a number of unique
and extraordinary properties, such as high charge carrier mobility, electronic energy spectrum without a gap between the conduction and valence bands, and frequency-independent absorption of EM radiation.
The optical properties of graphene have been extensively studied both
theoretically \cite{nmrPRB06,falkovsky,stauberBZ,stauberphonons,StauberGeim,carbotte,Juan,Mishchenko,rmp,rmpPeres,LiLouie,PRL,aires,APLPhotonic}
and experimentally \cite{nair,kuzmenko,mak,Crommieopt,kuzmenko2,kuzmenkoFaraday,NatureLoh,Li2008}.
Since the carrier concentration in graphene (and, hence, its frequency--dependent conductivity)
can be effectively tuned in wide limits by applying an external gate voltage \cite{Li2008}, it is a perspective material for tunable photonic
components. For example, in the area of plasmonics it is possible to make devices such as
tunable graphene-based switch\cite{BluVasPer2010}, polarizer\cite{Polarizer2012},
and polaritonic crystal\cite{PolCrys2012}. Moreover, using two\cite{Hwang2009,Svintsov2012,Stauber2012,PRIMER2013}
or more\cite{Eberlein2008,Jovanovic2011,Hajian2013,Metal2011,Sreekanth2012,Iorsh2012} parallel
 sheets of graphene can result in an unusual optical response of the structure owing to the interaction of charge carriers in the different layers by means of EM waves.
 Alternatively, for the same purposes it is possible to use an array of graphene ribbons\cite{Ju2011,nikitin_ribbon2012,Hipolito2012},
two-\cite{Yan_disks2012,Thongrattanasiri2012} or three-dimensional\cite{Berman2010,Yan2012}
arrays of graphene disks, or a two-dimensional array of antidots\cite{nikitin2012}.

The aim of the present work is twofold. On the one hand, in the studies considering
the transmittance of radiation through graphene\cite{LiLouie,aires,nair,kuzmenkoFaraday}
several authors analyzed only the case of normal incidence of the
radiation on the graphene sheet. By restraining themselves to this
particular case, these studies overlooked the unusual reflection and transmission properties taking place at oblique incidence. In this paper we discuss the transmission of EM
radiation through a graphene sheet when the impinging beam makes
an arbitrary angle, $\theta $, with the normal to the interface. We will show that, at grazing
incidence (i.e. for $\theta $ close to $90^{o}$) and when the
graphene layer is cladded by two identical dielectrics, it behaves like a mirror for $s$-polarized waves and is almost transparent for
$p$-polarized waves. (In contrast, if the dielectric constants of the media below and above the graphene sheet are
different, the interface reflects almost totally both $s$- and $p$-polarised waves as it is usual at grazing incidence).
Furthermore, for $\theta $ close to the total internal reflection angle, a single sheet of monolayer graphene strongly absorbs $s$-polarized waves, while there is almost no $p$-polarized absorption in these conditions.
On the other hand, we theoretically investigate the reflection of EM radiation, in the THz to far-infrared (FIR) range and for arbitrary $\theta$, from a periodic stack of parallel graphene sheets which constitute a semi-infinite one-dimensional (1D) photonic crystal. As it will be shown, this PC is highly reflective within certain frequency intervals corresponding to the gaps in its spectrum, and the widths of these gaps can be tuned by varying the gate
voltage giving the possibility to create a tunable mirror. Moreover,
when the angle of incidence for a $p$-polarized wave exceeds that of total internal reflection, it is possible to excite a surface EM mode supported by the semi-infinite photonic crystal.

\section{Single-layer graphene\label{sec:Single-layer-graphene}}

\begin{figure}[ht]
\begin{centering}
\includegraphics[clip,width=8.5cm]{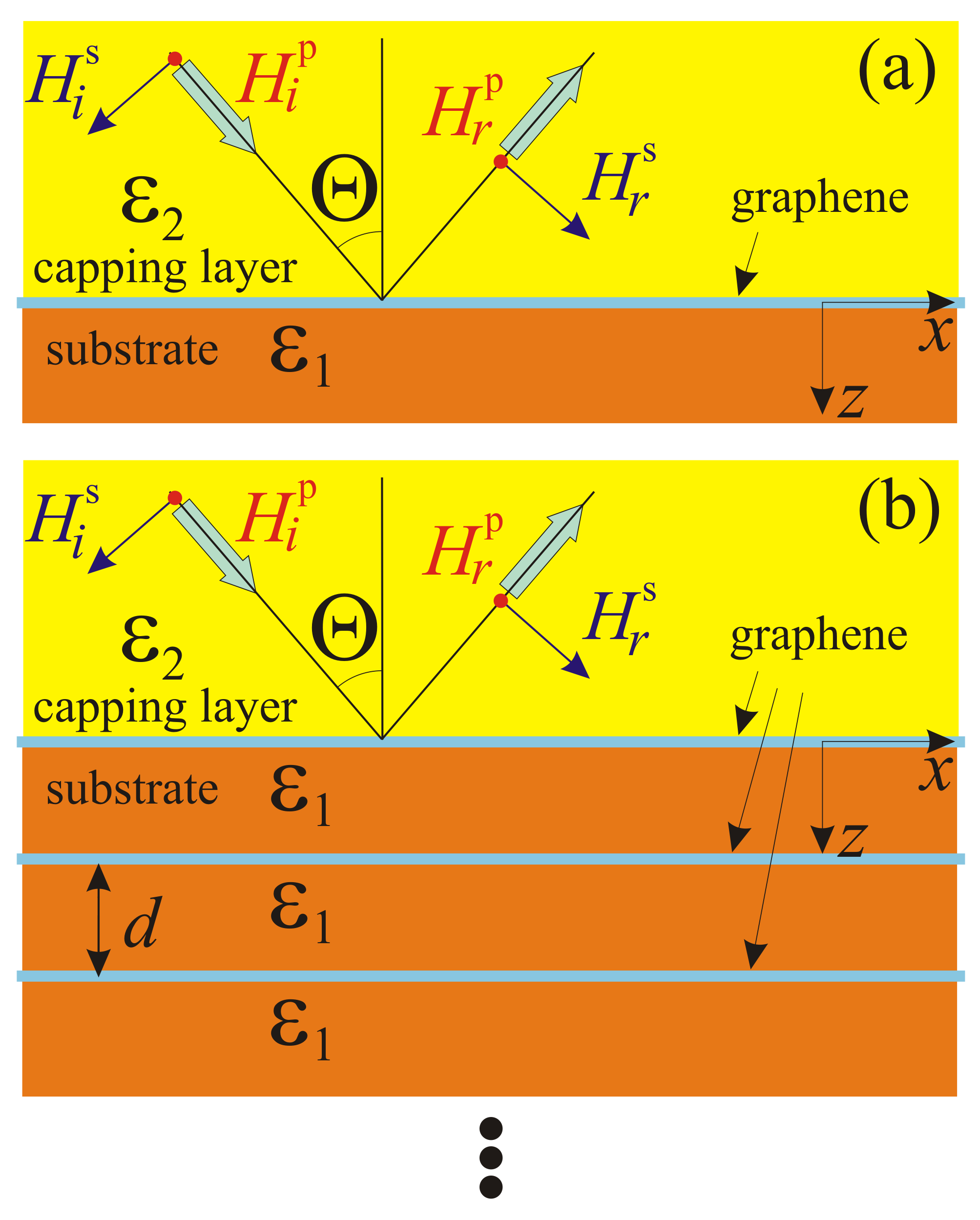}
\par\end{centering}

\caption{Schematic representation of the systems considered in Sec. \ref{sec:Single-layer-graphene}) (a) and \ref{sec:Graphene-multilayer-PC} (b), showing $p-$ or $s-$polarized incident and reflected waves.}
\label{fig:scheme}
\end{figure}

Let us first consider a single flat sheet of monolayer graphene located at the plane $z=0$ (so, the $z$ axis is perpendicular to it) and cladded by two semi-infinite dielectrics, a
substrate with a dielectric permittivity $\varepsilon_{1}>0$ and a capping medium with $\varepsilon_{2}>0$, occupying the half-spaces $z>0$
and $z<0$, respectively [see Fig. \ref{fig:scheme}(a)]. If the EM field is uniform along
the $y$ direction ($\partial/\partial y\equiv0$), it can be
decomposed into two separate waves with different polarizations.
Thus, a $p$-polarized (or TM) wave with the magnetic field perpendicular
to the plane of incidence ($xz$), possesses the electromagnetic field
components $\vec{E}=\left\{ E_{x},0,E_{z}\right\} ,$ $\vec{H}=\left\{ 0,H_{y},0\right\} $, while an $s$-polarized  (or TE) wave is described by electromagnetic
field components $\vec{E}=\left\{ 0,E_{y},0\right\} ,$ $\vec{H}=\left\{ H_{x},0,H_{z}\right\} $, with the electric field perpendicular to the plane
of incidence.

The temporal dependence of the fields is assumed of the form, $\vec{E}^{(j)},\vec{H}^{(j)}\sim\exp(-i\omega t)$,
where $\omega$ is the angular frequency of the radiation and the superscripts $j=1,2$
correspond to the electromagnetic field in the substrate and the capping dielectric, respectively.
Maxwell equations written explicitly for TE and TM waves in this particular situation can be found in several previous publications\cite{BluVasPer2010,Polarizer2012,PolCrys2012,PRIMER2013}.
We reckon that the EM wave falls on the interface $z=0$ from the capping dielectric side. In this case the
solution of the Maxwell equations for a $p$-polarized wave can be written as follows:
\begin{eqnarray}
H_{y}^{(2)}(x,z)=\left[H_{i}^{p}\exp(ik_{2,z}z)+H_{r}^{p}\exp(-ik_{2,z}z)\right]\times\nonumber \\
\exp(ik_{x}x),\label{eq:hrp}\qquad \\
H_{y}^{(1)}(x,z)=H_{t}^{p}\exp(ik_{x}x+ik_{1,z}z),\qquad \label{eq:htp}\\
E_{x}^{(2)}(x,z)=\frac{k_{2,z}}{\kappa\varepsilon_{2}}[H_{i}^{p}\exp(ik_{2,z}z)-H_{r}^{p}\exp(-ik_{2,z}z)]\times\nonumber \\
\exp(ik_{x}x),\qquad \label{eq:erp}\\
E_{x}^{(1)}(x,z)=\frac{k_{1,z}}{\kappa\varepsilon_{1}}H_{t}^{p}\exp(ik_{x}x+ik_{1,z}z),\qquad \label{eq:etp}
\end{eqnarray}
where
\begin{eqnarray}
k_{j,z}=\left(\kappa^{2}\varepsilon_{j}-k_{x}^{2}\right)^{1/2},\nonumber \\
k_{x}=\kappa\sqrt{\varepsilon_{2}}\sin\theta\label{eq:Kx}\:,
\end{eqnarray}
$\kappa=\omega/c$, $c$ is the velocity of light in vacuum.

At the same time, for an $s$-polarized wave we have:
\begin{eqnarray}
E_{y}^{(2)}(x,z)=\left[E_{i}^{s}\exp(ik_{2,z}z)+E_{r}^{s}\exp(-ik_{2,z}z)\right]\times\nonumber \\
\exp(ik_{x}x),\qquad \label{eq:ers}\\
E_{y}^{(1)}(x,z)=E_{t}^{s}\exp(ik_{x}x+ik_{1,z}z),\qquad \label{eq:ets}\\
H_{x}^{(2)}(x,z)=-\frac{k_{2,z}}{\kappa}[E_{i}^{s}\exp(ik_{2,z}z)-E_{r}^{s}\exp(-ik_{2,z}z)]\times\nonumber \\
\exp(ik_{x}x),\qquad \label{eq:hrs}\\
H_{x}^{(1)}(z)=-\frac{k_{1,z}}{\kappa}E_{t}^{s}\exp(ik_{x}x+ik_{1,z}z).\qquad \label{eq:hts}
\end{eqnarray}
In Eqs. (\ref{eq:hrp})--(\ref{eq:etp}) and (\ref{eq:ers})--(\ref{eq:hts}), $H_{i}^{p} \left (E_{i}^{s}\right )$, $H_{r}^{p} \left (E_{r}^{s}\right )$ and $H_{t}^{p} \left (E_{t}^{s}\right )$ denote the amplitudes of the incident, reflected and transmitted $p$- ($s$)-polarized waves, respectively.

In order to find the transmittance and the reflectance of the structure we apply boundary conditions at $z=0$, which include the continuity of the tangential
component of the electric field, $E_{x}^{(2)}(x,0)=E_{x}^{(1)}(x,0)$; $E_{y}^{(2)}(x,0)=E_{y}^{(1)}(x,0)$, and the discontinuity of the tangential component of the magnetic field caused by the induced surface currents in graphene,
$H_{y}^{(1)}(x,0)-H_{y}^{(2)}(x,0)=-(4\pi/c)j_{x}=-(4\pi/c)\sigma_{g}E_{x}(x,0)$,
$H_{x}^{(1)}(x,0)-H_{x}^{(2)}(x,0)=(4\pi/c)j_{y}=(4\pi/c)\sigma_{g}E_{y}(x,0)$,
where $\sigma_{g}$ is the graphene conductivity.
Matching the solutions for $z<0$ and $z>0$ using these boundary conditions, we obtain the amplitudes of the
reflected and transmitted waves,
\begin{eqnarray}
H_{r}^{p}=\frac{\varepsilon_{1}k_{2,z}-\varepsilon_{2}k_{1,z}+\frac{4\pi}{\omega}\sigma_{g}k_{2,z}k_{1,z}}{\varepsilon_{1}k_{2,z}+\varepsilon_{2}k_{1,z}+\frac{4\pi}{\omega}\sigma_{g}k_{2,z}k_{1,z}}H_{i}^{p},\label{eq:hr}\\
H_{t}^{p}=\frac{2\varepsilon_{1}k_{2,z}H_{i}^{p}}{\varepsilon_{1}k_{2,z}+\varepsilon_{2}k_{1,z}+\frac{4\pi}{\omega}\sigma_{g}k_{2,z}k_{1,z}},\label{eq:ht}
\end{eqnarray}
for $p$-polarization and
\begin{eqnarray}
E_{r}^{s}=-\frac{k_{1,z}-k_{2,z}+\frac{4\pi\omega}{c^{2}}\sigma_{g}}{k_{1,z}+k_{2,z}+\frac{4\pi\omega}{c^{2}}\sigma_{g}}E_{i}^{s}\label{eq:er}\\
E_{t}^{s}=\frac{2k_{2,z}E_{i}^{s}}{k_{1,z}+k_{2,z}+\frac{4\pi\omega}{c^{2}}\sigma_{g}}.\label{eq:et}
\end{eqnarray}
for $s$-polarization. The transmittance (reflectance) is expressed as the ratio of the Poynting vector $z$-components of the transmitted (reflected) and the incident
waves,
\begin{eqnarray*}
R_{p}=\left|\frac{H_{r}^{p}}{H_{i}^{p}}\right|^{2},\qquad T_{p}=\frac{k_{1,z}\varepsilon_{2}}{k_{2,z}\varepsilon_{1}}\left|\frac{H_{t}^{p}}{H_{i}^{p}}\right|^{2},\\
R_{s}=\left|\frac{E_{r}^{s}}{E_{i}^{s}}\right|^{2},\qquad T_{s}=\frac{k_{1,z}}{k_{2,z}}\left|\frac{E_{t}^{s}}{E_{i}^{s}}\right|^{2}.
\end{eqnarray*}

\begin{figure}[ht]
\begin{centering}
\includegraphics[clip,width=7cm]{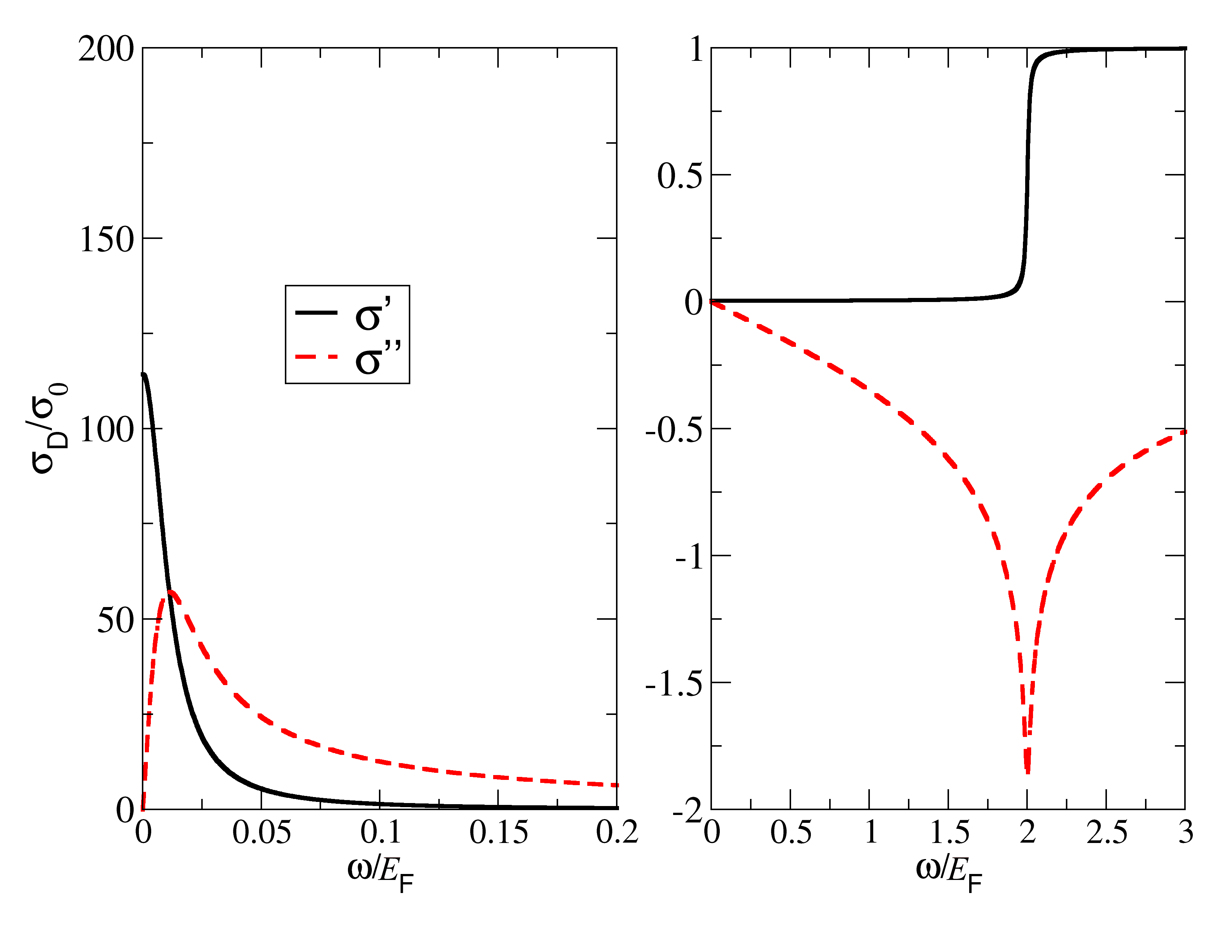}
\par\end{centering}

\caption{Optical conductivity of uniform graphene: Drude (left) and inter-band
(right) contributions. We assumed $E_{F}=0.23$ eV and $\Gamma=2.6$
meV. The solid (dashed) line stands for the real (imaginary) part
of the conductivity.}
\label{fig_conductivity}
\end{figure}

Since $R$ and $T$ are determined by the conductivity $\sigma_{g}$ {[}see Eqs. (\ref{eq:hr})--(\ref{eq:et}){]},
we briefly consider its frequency dependence. The frequency--dependent (optical) conductivity of graphene is a sum of two contributions: (i) a Drude term describing
intra-band processes, and (ii) a term taking into account inter-band transitions.
At zero temperature the optical conductivity has a simple analytical
expression\cite{nmrPRB06,falkovsky,rmp,rmpPeres,StauberGeim}.  The
inter-band contribution has the form $\sigma_{I}=\sigma_{I}'+i\sigma_{I}''$,
where
\begin{eqnarray}
\sigma_{I}'=\sigma_{0}\left(1+\frac{1}{\pi}\arctan\frac{\hbar\omega-2E_{F}}{\hbar\Gamma}\right.\nonumber \\
-\left.\frac{1}{\pi}\arctan\frac{\hbar\omega+2E_{F}}{\hbar\Gamma}\right)\,,
\end{eqnarray}
and
\begin{equation}
\sigma_{I}''=-\sigma_{0}\frac{1}{2\pi}\ln\frac{(2E_{F}+\hbar\omega)^{2}+\hbar^{2}\Gamma^{2}}{(2E_{F}-\hbar\omega)^{2}+\hbar^{2}\Gamma^{2}}\,,
\end{equation}
where $\sigma_{0}=\pi e^{2}/(2h)$ is the so called universal conductivity
of graphene. The Drude conductivity term is
\begin{equation}
\sigma_{D}=\sigma_{0}\frac{4E_{F}}{\pi}\frac{1}{\hbar\Gamma-i\hbar\omega}\,,\label{eq_sigma_xx_semiclass}
\end{equation}
where $\Gamma$ is the inverse of the momentum relaxation time and $E_{F}>0$ is the Fermi level
position with respect to the Dirac point. The total conductivity is
\begin{equation}
\sigma_{g}=\sigma_{I}'+i\sigma_{I}''+\sigma_{D}\,.
\end{equation}
We can write $\sigma_{g}=\sigma_{0}f(\omega)$, where $f(\omega)$
is a dimensionless function. In Fig. \ref{fig_conductivity} we depict
the Drude and inter-band contributions to the total optical conductivity
of graphene.

\begin{figure}[htb]
\begin{centering}
\includegraphics[clip,width=8cm]{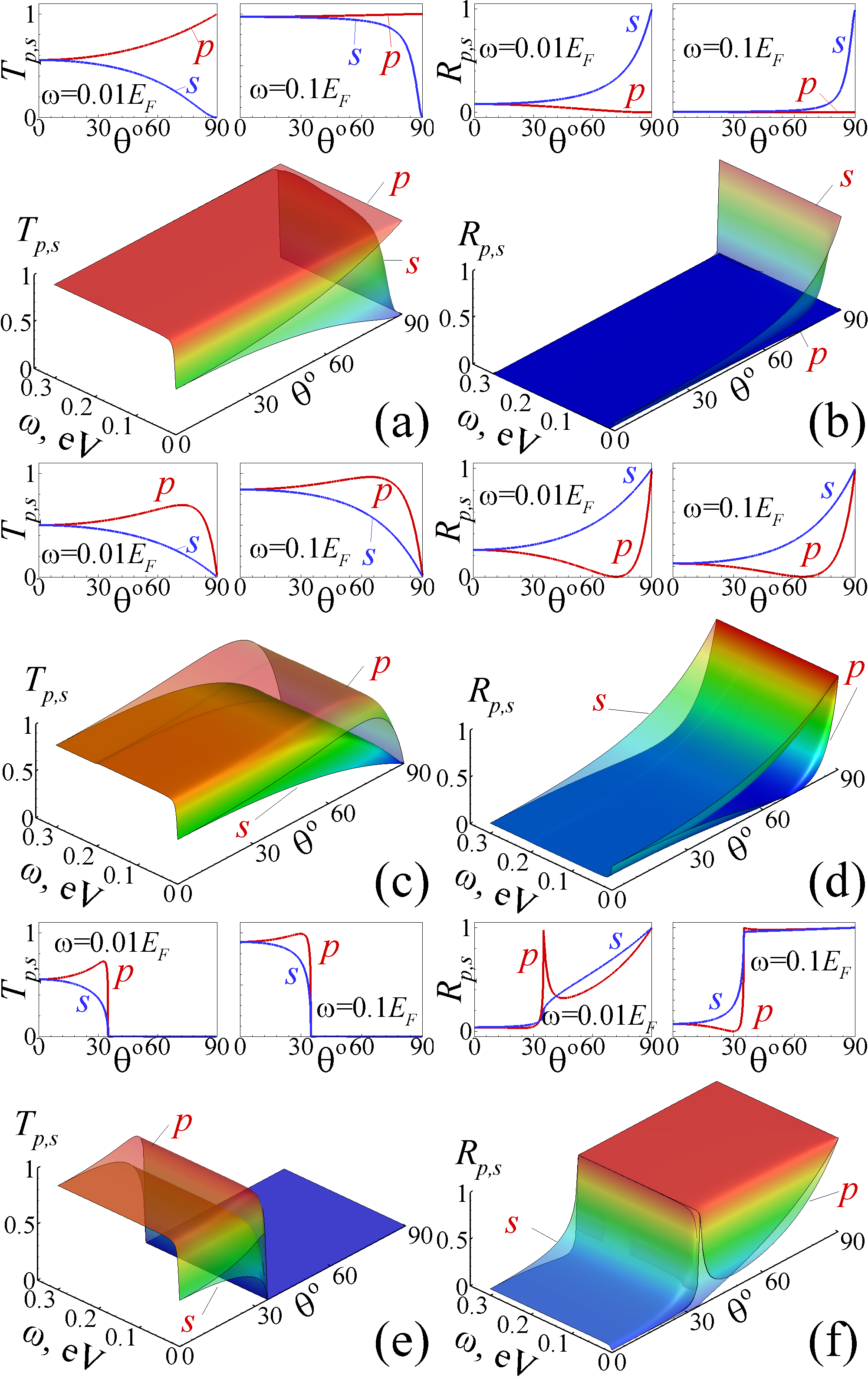}
\par\end{centering}

\caption{Transmittance $T_{p,s}$ (left column) and reflectance $R_{p,s}$
(right column) of graphene cladded by two semi--infinite dielectrics {\it versus} angle of incidence $\theta$
and frequency $\omega$. In all cases $\varepsilon_{1}=3.9$ (SiO$_2$) and $\Gamma=2.6\,$meV. Other parameters are: $\varepsilon_{2}=3.9$, $E_{F}=0.157\,$eV (upper
row), $\varepsilon_{2}=1.0$, $E_{F}=0.1\,$eV (middle row), or $\varepsilon_{2}=11.9$ (Si), $E_{F}=0.25\,$eV (lower row). In each panel, the angular dependences for two fixed frequencies ($\hbar \omega=0.01E_F$ and $\hbar \omega=0.1E_F$) are depicted in the insets.}

\label{fig:transmittance}
\end{figure}

\begin{figure}[htb]
\begin{centering}
\includegraphics[clip,width=8cm]{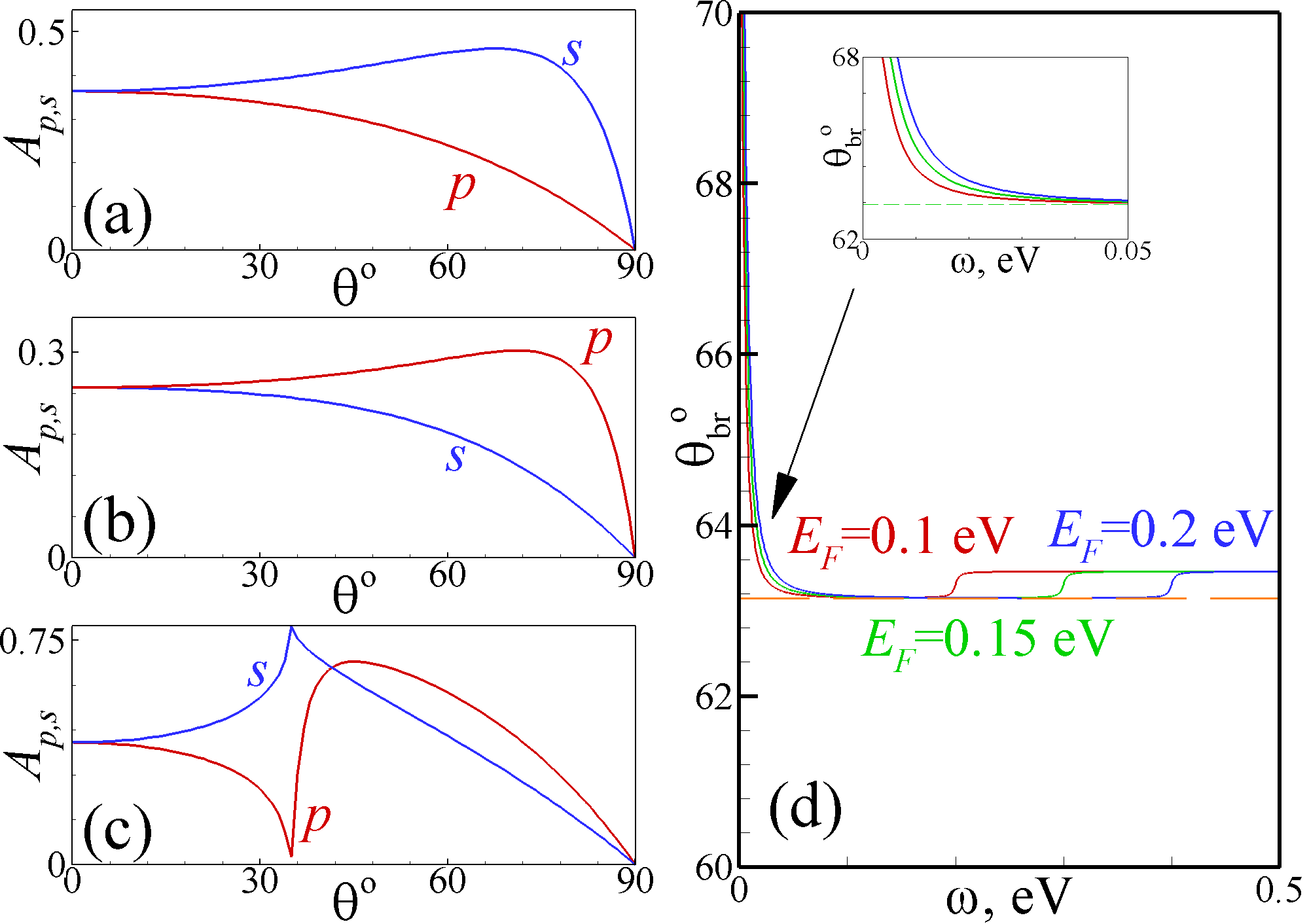}
\par\end{centering}

\caption{(a-c) Absorbance, $A_{p,s}$, at frequency $\omega=0.01E_{F}$ for $p-$ and $s-$polarizations as function of the
angle of incidence. The parameters of panels (a), (b), and (c) are
the same as the same as for the upper, middle, and lower rows in Fig. \ref{fig:transmittance}, respectively. (d) Frequency dependence of the \textit{quasi}-Brewster angle, $\theta_{br}$, for $\varepsilon_{2}=1$, $\varepsilon_{1}=3.9$
and three different values of $E_F$ as marked on the plot. The dashed horizontal line depicts the conventional Brewster angle $\theta_{br}^0$ of the interface
without graphene. The inset shows the low-frequency region magnified.}
\label{fig:absorbance}
\end{figure}

It is evident that at low-frequencies (left panel in Fig. \ref{fig_conductivity})
the Drude term significantly exceeds the interband one (both real
and imaginary parts), while in the high-frequency range (right panel in
Fig. \ref{fig_conductivity}) the interband
term dominates. Moreover, in the vicinity of the threshold frequency,
$\omega=2E_{F}/\hbar$, the real part of the conductivity increases drastically
and achieves the universal value, $\sigma_{0}$ (onset of interband transitions), while the imaginary
part is minimal, negative and of the order of several universal conductivities in modulus.
As a result, at low frequencies the presence of graphene at the interface
between two dielectrics influences significantly the reflectance
and the transmittance of the structure.
This effect, owing to the high value of the Drude conductivity, is clearly seen in Figs. \ref{fig:transmittance}
and \ref{fig:absorbance}(a--c), where  the low-frequency
region is characterized by the lower transmittance {[}see Figs. \ref{fig:transmittance}(a),
\ref{fig:transmittance}(c), \ref{fig:transmittance}(e) and the respective insets for $\omega=0.01E_F${]}, higher
reflectance {[}Figs. \ref{fig:transmittance}(b), \ref{fig:transmittance}(d), \ref{fig:transmittance}(f) and the respective insets for $\omega=0.01E_F${]} and enhanced absorbance {[}Figs. \ref{fig:absorbance}(a)--\ref{fig:absorbance}(c){]} for all parameters' values.
In fact, the transmittance and the reflectance are mainly determined by
the real part of the conductivity, except when the imaginary part
is large in modulus (at low frequencies and near $\hbar \omega=2E_{F}$).

At normal incidence ($\theta=0$), we have (for any polarization):
\begin{eqnarray}
T_{p}=T_{s}=\sqrt{\varepsilon_{2}\varepsilon_{1}}\left|\frac{2}{\sqrt{\varepsilon_{1}}+\sqrt{\varepsilon_{2}}+\pi\alpha f(\omega)}\right|^{2},\label{eq:Tzero}\\
R_{p}=R_{s}=\left|\frac{\sqrt{\varepsilon_{1}}-\sqrt{\varepsilon_{2}}+\pi\alpha f(\omega)}{\sqrt{\varepsilon_{1}}+\sqrt{\varepsilon_{2}}+\pi\alpha f(\omega)}\right|^{2}\:,
\label{eq:Rzero}
\end{eqnarray}
where $\alpha$ is the fine structure constant. For oblique incidence ($\theta\ne0$), the dependencies of the transmittance
and the reflectance on $\omega $ are strongly affected by the relation between the dielectric
permittivities of the substrate and the capping dielectric. Therefore, we considered all three possible situations: (i) $\varepsilon_{1}=\varepsilon_{2}$
{[}Figs. \ref{fig:transmittance}(a), \ref{fig:transmittance}(b){]}, (ii) $\varepsilon_{1}>\varepsilon_{2}$ {[}Figs. \ref{fig:transmittance}(c),
\ref{fig:transmittance}(d){]}, and (iii) $\varepsilon_{1}<\varepsilon_{2}$ {[}Figs. \ref{fig:transmittance}(e), \ref{fig:transmittance}(f){]}.
In the "symmetric" case of $\varepsilon_{1}=\varepsilon_{2}=\varepsilon$, the transmittance and the reflectance can be expressed by simple formulae,
\begin{eqnarray}
R_{p}=\left|\frac{\pi\alpha f(\omega)\cos\theta/\sqrt{\varepsilon }}{2+\pi\alpha f(\omega)\cos\theta/\sqrt{\varepsilon }}\right|^{2},\label{eq:rp}\\
T_{p}=\left|\frac{2}{2+\pi\alpha f(\omega)\cos\theta/\sqrt{\varepsilon }}\right|^{2},\\
R_{s}=\left|\frac{\pi\alpha f(\omega)/\sqrt{\varepsilon }}{2\cos\theta+\pi\alpha f(\omega)\cos\theta/\sqrt{\varepsilon }}\right|^{2},\\
T_{s}=\left|\frac{2\cos\theta}{2\cos\theta+\pi\alpha f(\omega)/\sqrt{\varepsilon }}\right|^{2}\!.\label{eq:ts}
\end{eqnarray}
Note that the factor $4\pi\sigma_{0}/c=\pi\alpha$ multiplying the dimensionless function $f(\omega )$, which represents the frequency dependence of the graphene conductivity, is a small number (=0.023). Thus, unless the absolute value of $f(\omega )$ is large, the term related to graphene in Eqs. (\ref{eq:hr})--(\ref{eq:et}) and, accordingly, in the above expressions for $T$ and $R$ is small. Therefore, the reflectance and the transmittance of the structure are close to the values defined by usual Fresnel's expressions, except for $\omega \rightarrow 0$ and $\omega \approx 2E_F/\hbar $. In particular, the reflectance is proportional to $(\pi\alpha )^2$.

It should also be noticed that in Eqs.(\ref{eq:Tzero}), (\ref{eq:Rzero}) and (\ref{eq:rp})--(\ref{eq:ts}) the effect of graphene is stronger for lower dielectric constants and is maximal for free standing graphene ($\varepsilon_{1,2}=1$).
As $\theta$ increases {[}see Figs. \ref{fig:transmittance}(a), \ref{fig:transmittance}(b) and the insets{]}, the transmittance $T_{s}$
decreases and attains zero for $\theta=\pi/2$, while the reflectance
$R_{s}$ increases and tends to unity at $\theta \rightarrow \pi/2$. In contrast, a $p$-polarized wave is "totally transmitted"
at $\theta \rightarrow \pi/2$ ($R_{p}=0$, $T_{p} \rightarrow 1$).
With the electric field perpendicular to the graphene sheet (TM wave), no charge oscillations are induced at the interface and the EM field is not perturbed.
Also, the low-frequency absorbance at grazing incidence  is a decreasing function of the angle for both $p$- and $s$-polarized waves and the limit $\theta=\pi/2$ corresponds to zero absorbance {[}see Fig. \ref{fig:absorbance}(a){]}.
Note that the absorption is entirely related to graphene because the dielectrics are assumed dispersionless.

The situation is different when the dielectric constants of the substrate and capping dielectric are not equal because in this case there would be reflection even in the absence of graphene.
Here, at grazing incidence the reflectance (transmittance) is close to unity(zero) for \textit{both} polarizations {[}see Figs. \ref{fig:transmittance}(c)--\ref{fig:transmittance}(f){]}, just like for a "normal" interface between two dielectrics (without graphene). In the case (ii), although the angular dependence of the low-frequency absorbance {[}Fig. \ref{fig:absorbance}(b){]}
is qualitatively similar to the "symmetric" case (i), it is higher for TM waves {[}cf. Fig. \ref{fig:absorbance}(b)
and Fig. \ref{fig:absorbance}(a){]}, in contrast with the case (i).
The particularity of the case (ii) ($\varepsilon_{1}>\varepsilon_{2}$, e.g. uncovered graphene on a substrate) is that the angular dependence
of the TM-wave reflectance {[}see Fig. \ref{fig:transmittance}(d){]} possesses
a minimum at a certain $\theta$ close to the Brewster angle for the two dielectrics considered, $\theta_{br}^0=\mbox {arctg}[(\varepsilon_{1}/\varepsilon_{2})^{1/2}]$\cite{Born-Wolf}.
Owing to the non-zero imaginary part of the graphene conductivity, the TM wave reflectance at this angle is finite, while it would reach zero in the case without graphene. Therefore we call this $\theta$ the  ``\textit{quasi}-Brewster'' angle, $\theta_{br}$. It depends upon the frequency and the Fermi energy of graphene; this dependence
is depicted in Fig. \ref{fig:absorbance}(d). At low frequencies, where the conductivity $\sigma_{g}$ is high owing to the Drude term, the quasi-Brewster
angle of the structure exceeds the conventional Brewster angle $\theta_{br}^0$ of the interface without graphene. This effect is more pronounced for
higher values of the Fermi energy -- compare the three curves in the plot.
When the frequency grows (and the Drude term in $\sigma_{g}$ becomes smaller) $\theta_{br}$ approaches the value
of the conventional Brewster angle $\theta_{br}^0$. However, at $\omega\sim 2E_{F}$ the quasi-Brewster angle jumps up because of the Fermi step in the real part of $\sigma_{g}$ (onset of interband transitions).
The difference between $\theta_{br}$ and the conventional Brewster angle, $\theta_{br}^0$, characteristic of graphene--free interface between the same two dielectrics, can be used for visualization of graphene, which constitutes a considerable problem \cite {Schellenberg}. At low frequencies, $\theta_{br}-\theta_{br}^0$ can reach easily detectable values of $\sim 5^o$.

In the last possible situation (iii), there is a critical angle of total internal reflection [$\approx35^{o}$ for the parameters of Figs. \ref{fig:transmittance}(e) and \ref{fig:transmittance}(f)]. Above
this critical angle, $\theta _c$, the transmittance vanishes and the reflectance is close to unity {[}see Figs. \ref{fig:transmittance}(e) and \ref{fig:transmittance}(f){]}.
Notice that the value of $\theta _c$ does not depend upon the graphene parameters.
In the case under consideration,
the low-frequency absorbance {[}Fig. \ref{fig:absorbance}(c){]} exhibits the most interesting behavior: in the vicinity of the total internal
reflection angle the absorbance of the $p-$polarized wave is almost zero, while
that of the $s-$polarized wave reaches its maximum of $\approx75\%$.

\section{Graphene multilayer photonic crystal\label{sec:Graphene-multilayer-PC}}

Now we shall consider an external EM wave falling on the periodic multilayer structure [see Fig.~\ref{fig:scheme}(b)] consisting of an infinite number of parallel monolayer graphene sheets separated by dielectric slabs of thickness
$d$; in practical terms few graphene layers play the same role as an infinite number of them \cite{EPL2013}.
The geometry of the problem is similar to considered in Sec.II, but with graphene layers (for which we shall assume the same Fermi energy) located at positions $z=md$, $m\in\left[0,\infty\right)$. Thus,
the considered structure is a semi-infinite 1D PC, terminated by the graphene layer at $z=0$. We shall still consider a capping dielectric (with arbitrary real $\varepsilon_{2}$) on top of it.

In order to find the reflectance of this PC, we notice, that the structure
of the electromagnetic field in the capping dielectric is the same, as
represented by Eqs. (\ref{eq:hrp}),(\ref{eq:erp}) and (\ref{eq:ers}),(\ref{eq:hrs})
for $p$- and $s$-polarized waves, respectively. At the same time, the fields in
the substrate should be considered separately in each layer
between adjacent graphene sheets at planes $z=md$ and $z=(m+1)d$.
Namely, solutions of the Maxwell equations at spatial domain $md\leq z\leq(m+1)d$ can be represented as
\begin{eqnarray}
H_{y}^{(1)}(x,z)=\left\{ H_{+}^{(m)}\exp\left[ik_{1,z}\left(z-md\right)\right]+\right.\label{eq:htp-sl}\\
\left.H_{-}^{(m)}\exp\left[-ik_{1,z}\left(z-md\right)\right]\right\} \exp(ik_{x}x),\nonumber \\
E_{x}^{(1)}(x,z)=\frac{k_{1,z}}{\kappa\varepsilon_{1}}\left\{ H_{+}^{(m)}\exp\left[ik_{1,z}\left(z-md\right)\right]-\right.\label{eq:etp-sl}\\
\left.H_{-}^{(m)}\exp\left[-ik_{1,z}\left(z-md\right)\right]\right\} \exp(ik_{x}x)\nonumber
\end{eqnarray}
and
\begin{eqnarray}
E_{y}^{(1)}(x,z)=\left\{ E_{+}^{(m)}\exp\left[ik_{1,z}\left(z-md\right)\right]+\right.\label{eq:ets-sl}\\
\left.E_{-}^{(m)}\exp\left[-ik_{1,z}\left(z-md\right)\right]\right\} \exp(ik_{x}x),\nonumber \\
H_{x}^{(1)}(x,z)=-\frac{k_{1,z}}{\kappa}\left\{ E_{+}^{(m)}\exp\left[ik_{1,z}\left(z-md\right)\right]-\right.\label{eq:hts-sl}\\
\left.E_{-}^{(m)}\exp\left[-ik_{1,z}\left(z-md\right)\right]\right\} \exp(ik_{x}x).\nonumber
\end{eqnarray}
Here $H_{\pm}^{(m)}$ are the amplitudes for forward (sign ''+'')
or backward (sign \textquotedblright{}--\textquotedblright{}) propagating
TM waves. Correspondingly, $E_{\pm}^{(m)}$ represent the amplitudes
of the TE waves. The amplitudes $H_{\pm}^{(m+1)}$ can be related to $H_{\pm}^{(m)}$
by matching boundary conditions at $z=(m+1)d$ on graphene (similar
to that, used in Sec. \ref{sec:Single-layer-graphene}), namely:
\begin{eqnarray}
\left(\begin{array}{c}
H_{+}^{(m+1)}\\
H_{-}^{(m+1)}
\end{array}\right)=\hat{M}_{p}\left(\begin{array}{c}
H_{+}^{(m)}\\
H_{-}^{(m)}
\end{array}\right),\label{eq:h-mp}\\
\hat{M}_{p}=\left(\begin{array}{cc}
e^{ik_{1,z}d}\left[1-\frac{2\pi k_{1,z}}{\omega\varepsilon_{1}}\sigma_{g}\right] & e^{-ik_{1,z}d}\frac{2\pi k_{1,z}}{\omega\varepsilon_{1}}\sigma_{g}\\
-e^{ik_{1,z}d}\frac{2\pi k_{1,z}}{\omega\varepsilon_{1}}\sigma_{g} & e^{-ik_{1,z}d}\left[1+\frac{2\pi k_{1,z}}{\omega\varepsilon_{1}}\sigma_{g}\right]
\end{array}\right).\nonumber
\end{eqnarray}
Similarly, for $s-$polarization,
\begin{eqnarray}
\left(\begin{array}{c}
E_{+}^{(m+1)}\\
E_{-}^{(m+1)}
\end{array}\right)=\hat{M}_{s}\left(\begin{array}{c}
E_{+}^{(m)}\\
E_{-}^{(m)}
\end{array}\right),\label{eq:e-ms}\\
\hat{M}_{s}=\left(\begin{array}{cc}
e^{ik_{1,z}d}\left[1-\frac{2\pi\omega}{c{}^{2}k_{1,z}}\sigma_{g}\right] & -e^{-ik_{1,z}d}\frac{2\pi\omega}{c{}^{2}k_{1,z}}\sigma_{g}\\
e^{ik_{1,z}d}\frac{2\pi\omega}{c{}^{2}k_{1,z}}\sigma_{g} & e^{-ik_{1,z}d}\left[1+\frac{2\pi\omega}{c{}^{2}k_{1,z}}\sigma_{g}\right]
\end{array}\right).\nonumber
\end{eqnarray}
Since the considered structure is periodic \footnote {The surface at $z=0$, of course, breaks the translational symmetry. This symmetry breaking can give rise to local (surface) modes, known as Tamm states in the case of electronic structure of crystals (not to be confused with the Bloch-type TM surface wave discussed below!). However, the band structure of the spectrum is preserved and the surface EM mode can be excited only under special conditions, enhancing the in-plane wavevector \cite{BluVasPer2010,PRIMER2013}.}, it is possible to use the
Bloch theorem, which determines the proportionality between the field
amplitudes in the adjacent periods through the Bloch wavevector $q$:
\begin{eqnarray*}
H_{\pm}^{(m+1)} & = & \exp\left(iqd\right)H_{\pm}^{(m)},\\
E_{\pm}^{(m+1)} & = & \exp\left(iqd\right)E_{\pm}^{(m)}.
\end{eqnarray*}
After substitution of these relations into Eqs.(\ref{eq:h-mp}),(\ref{eq:e-ms}),
the compatibility condition of the resulting linear equations requires:
 \begin{eqnarray}
\mathrm{Det}\left|\hat{M}_{p}-\exp\left(iqd\right)\hat{I}\right|=0,\label{eq:disp-p-mat}\\
\mathrm{Det}\left|\hat{M}_{s}-\exp\left(iqd\right)\hat{I}\right|=0,\label{eq:disp-s-mat}
\end{eqnarray}
where $\hat{I}$ is the unity matrix.
Equations (\ref {eq:disp-p-mat}, \ref{eq:disp-s-mat}) yield the dispersion relations for $p-$ and $s-$polarized EM waves in the graphene multilayer
PC:
\begin{equation}
\cos\left(qd\right)-\cos\left(k_{1,z}d\right)+i\frac{2\pi k_{1,z}}{\omega\varepsilon_{1}}\sigma_{g}\sin\left(k_{1,z}d\right)=0\:,
\label{eq:disp-p}
\end{equation}
and
\begin{equation}
\cos\left(qd\right)-\cos\left(k_{1,z}d\right)+i\frac{2\pi\omega}{c^{2}k_{1,z}}\sigma_{g}\sin\left(k_{1,z}d\right)=0\:.
\label{eq:disp-s}
\end{equation}
We note that similar expressions have been obtained in Ref.\cite{Hajian2013}

\begin{figure}
\begin{centering}
\includegraphics[width=8.5cm]{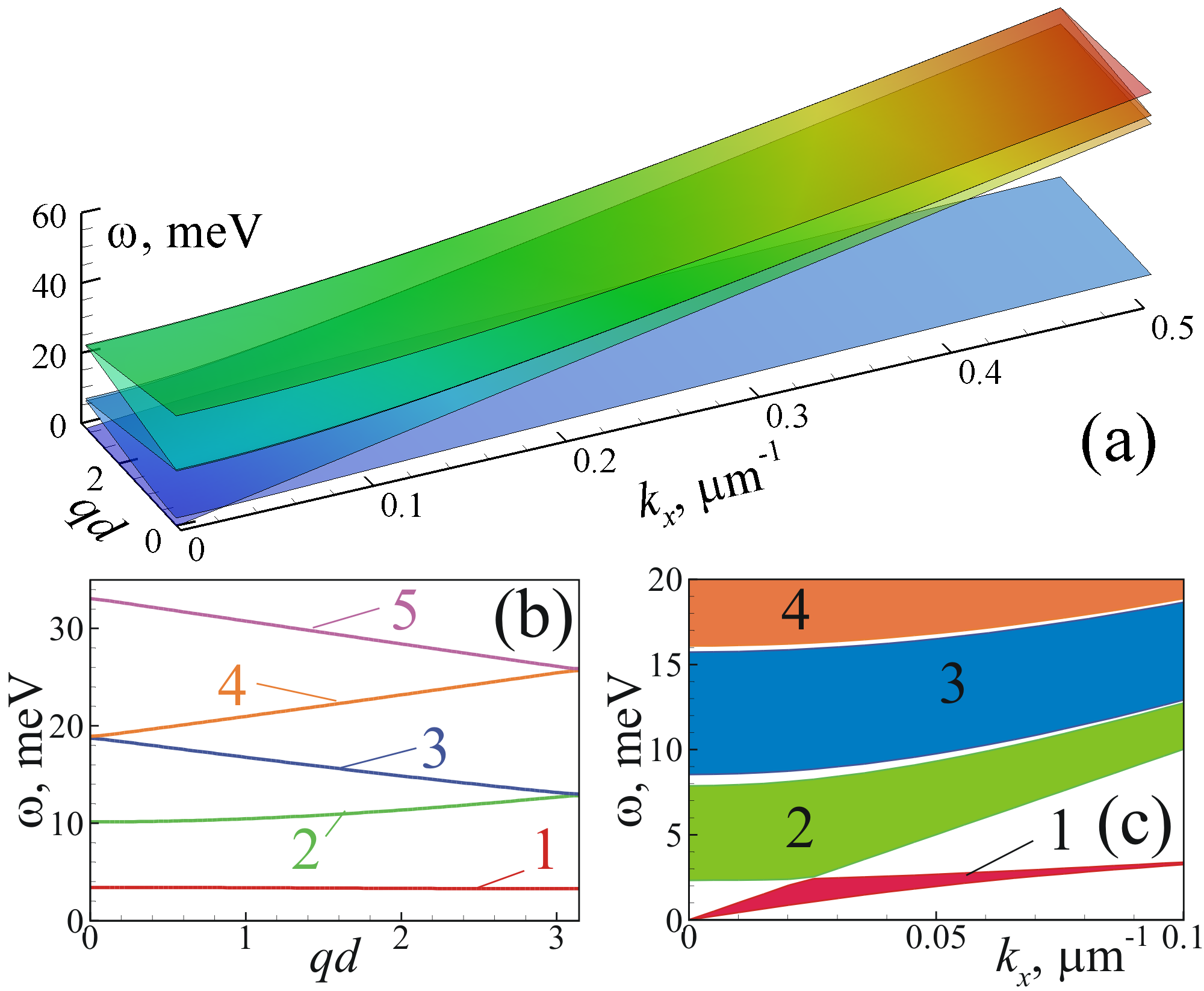}
\par\end{centering}

\caption{(a) TM wave eigenfrequencies \textit{versus} $x-$component
of the wavevector, $k_{x}$, and Bloch wavevector, $q$; (b) Eigenfrequencies
$\omega$ \textit{vs} $q$ for fixed $k_{x}=0.1\,\mu\mathrm{m}^{-1}$; (c) Eigenfrequencies
$\omega$ \textit{vs} $k_{x}$, dashed zones correspond to the allowed bands with the boundaries
determined by $q=0$ and $q=\pi/d$. Other parameters are: $d=40\,\mu$m, $E_{F}=0.157\,$eV, $\varepsilon_{1}=3.9$, $\Gamma=0$.
The numbers designate different allowed bands, 1 for surface mode, 2, 3, {\it etc} for bulk modes.
\label{fig:sl-p} }
\end{figure}

\begin{figure}
\begin{centering}
\includegraphics[width=8.5cm]{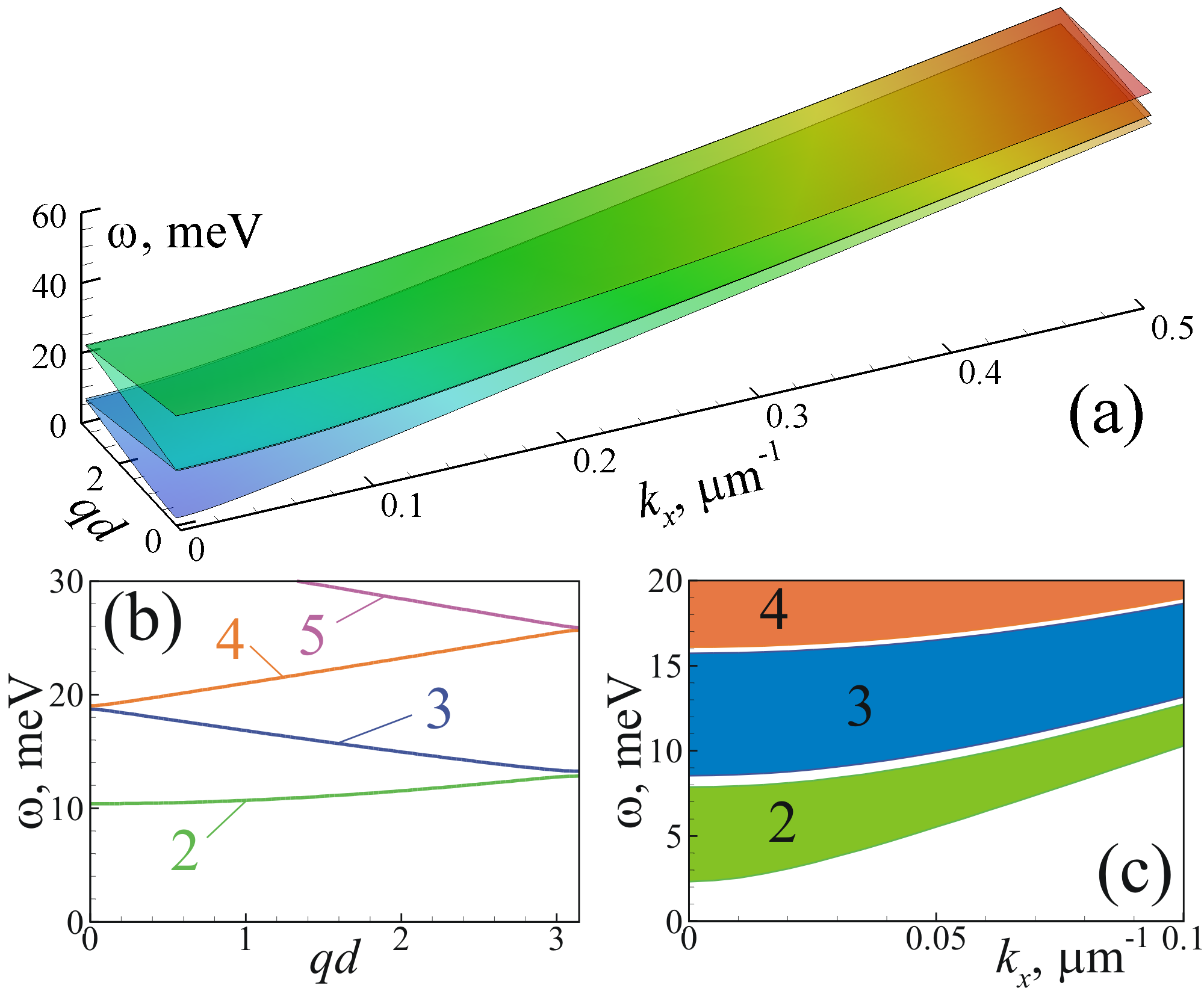}
\par\end{centering}

\caption{The same as in Fig. \ref{fig:sl-p}, but for TE waves. There is no allowed surface mode in this case.
\label{fig:sl-s}}
\end{figure}

Before considering the dispersion properties of $p-$ and $s-$polarized waves in detail,
it should be noticed, that dispersion curves depicted in Figs. \ref{fig:sl-p}
and \ref{fig:sl-s} have been calculated for zero damping, $\Gamma=0$,
when the graphene conductivity possesses only the imaginary part.
As a result, the eigenfrequencies, the in-plane component of the wavevector $k_{x}$, and the Bloch
wavevector $q$ are real values. In the case of nonzero $\Gamma$ the
eigenfrequencies will be complex values with imaginary part characterizing the mode damping.
The calculated spectra exhibit the band structure of a photonic crystal for both $p-$ (Fig. \ref{fig:sl-p}) and $s-$polarized waves
(Fig. \ref{fig:sl-s}).
In particular, there are gaps in the spectra {[}see Figs. \ref{fig:sl-p}(a) and \ref{fig:sl-s}(a){]} that appear in the center ($q=0$) and in the edges ($q=\pm \pi/d$)
of the first Brillouin zone, and the widths of gaps decrease with the increase
of $k_{x}$ {[}see Figs. \ref{fig:sl-p}(c)
and \ref{fig:sl-s}(c){]}. However, the main feature of the
$p-$polarization spectrum is the presence of \textit{surface} mode with purely
imaginary $k_{1,z}$ [marked by 1 (red color) in Figs. \ref{fig:sl-p}(b) and \ref{fig:sl-p}(c)]
together with \textit{bulk} modes with purely real $k_{1,z}$ {[}marked
by 2 (green), 3 (blue), 4 (orange) and 5 (pink) colors in Figs. \ref{fig:sl-p}(b) and \ref{fig:sl-p}(c){]}.
The former is a "Bloch surface plasmon-polariton", with the electric and magnetic fields strongly localized at the graphene sheets \cite{PRIMER2013} but with a \textit{real} Bloch wavevector.
In the case of $s$ polarization such a surface mode does not exist and only
bulk modes are present in the spectrum {[}see Figs. \ref{fig:sl-s}(b) and \ref{fig:sl-s}(c){]}.

\begin{figure}
\begin{centering}
\includegraphics[width=8.5cm]{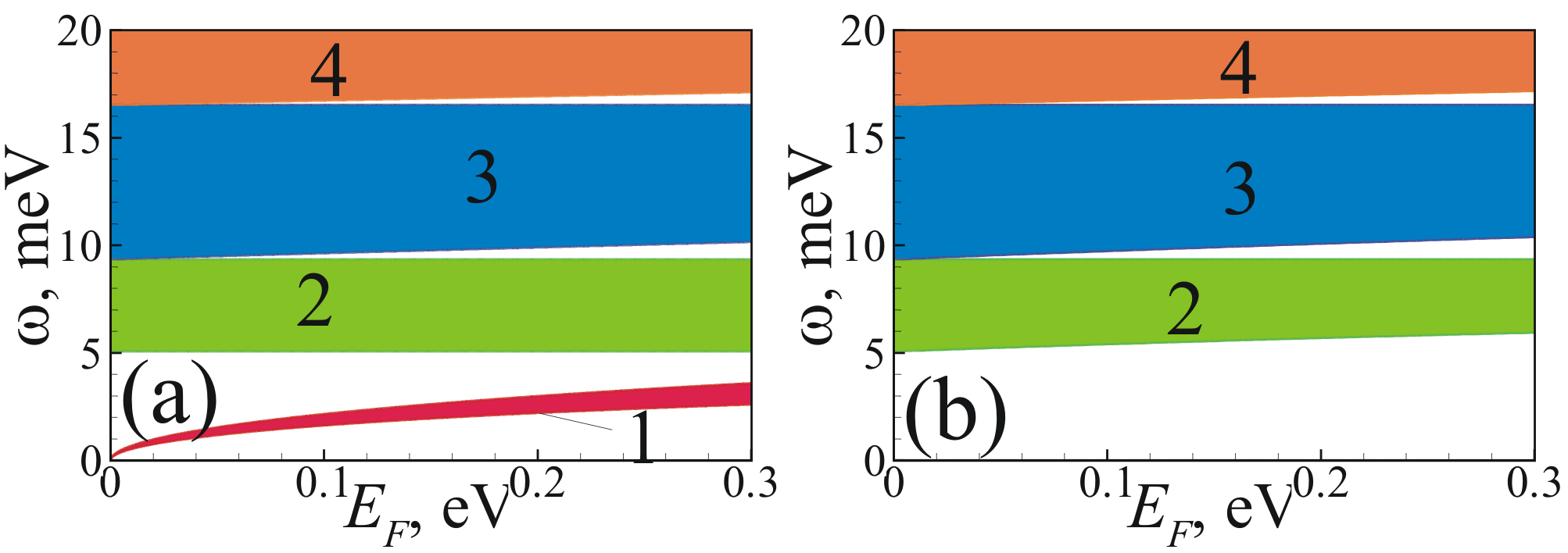}
\par\end{centering}
\caption{Eigenfrequencies for $p-$ (a) and $s-$polarized (b) waves \textit{vs} graphene Fermi energy for a fixed $k_{x}$($=0.05\,\mu\mathrm{m}^{-1}$). Other parameters are: $\varepsilon_{1}=3.9$, $\Gamma=0$, $d=40\,\mu$m. As in Figs. \ref{fig:sl-p} and \ref{fig:sl-s}, dashed zones correspond to allowed bands.
\label{fig:var-gap-eigen}}
\end{figure}

Particular solutions of the dispersion relation for $p$ polarization, Eq. (\ref{eq:disp-p}),
for $q=\pi/d$ {[}so called "Bragg modes"{]} can be represented
as
\begin{equation}
k_{1,z}=\left(\frac{\omega}{c^{2}}^{2}\varepsilon_{1}-k_{x}^{2}\right)^{1/2}=\left(2n+1\right)\pi/d,\qquad n\in\left[0,\infty\right).\label{eq:w-g-pi}
\end{equation}
For $q=0$ we have:
\begin{equation}
k_{1,z}=\left(\frac{\omega}{c^{2}}^{2}\varepsilon_{1}-k_{x}^{2}\right)^{1/2}=2n^{\prime}\pi/d,\qquad n^{\prime}\in\left[0,\infty\right).\label{eq:w-g-0}
\end{equation}
$S-$polarized Bragg modes {[}solutions of (\ref{eq:disp-s}){]} for $q=\pi/d$
are exactly the same as (\ref{eq:w-g-pi}), but for $q=0$ they are similar
to (\ref{eq:w-g-0}) (except that $n^{\prime}\neq 0$). The modes with Bragg wavevectors, (\ref{eq:w-g-pi}) and (\ref{eq:w-g-0}), have nodes at graphene layers and, therefore, these solutions do
not involve the graphene conductivity, $\sigma_g$. As a matter of fact,
these solutions correspond to $H_{+}^{(m)}=H_{-}^{(m)}$ for $p$ polarization and
$E_{+}^{(m)}=-E_{-}^{(m)}$ for $s$ polarization. It implies zero in-plane components of the electric field in both cases, consequently, no electric current is induced in graphene sheets located
at $z=md$ {[}see Eqs. (\ref{eq:etp-sl}) and (\ref{eq:ets-sl}){]}.

Secondly, in the case of $p$ polarization $k_{1,z}=0$ is the solution that implies
arbitrary $H_{+}^{(m)}$ and $H_{-}^{(m)}$, and, as a result, $H_{y}^{(1)}$
independent upon $z$ as well as $E_{x}^{(1)}\equiv0$. At the same
time, for $s$ polarization the solution $k_{1,z}=0$ corresponds to a trivial solution of the Maxwell equations with zero electric
and magnetic fields. For $p-$polarization, the line $k_{1,z}=0$ is crossed by another dispersion curve at the point $k_{x}=\sqrt {4\alpha E_{F}/(\hbar cd)}$
{[}see Fig. \ref{fig:sl-p}(c){]}, where there is
no gap between the surface and bulk mode bands. Below this point, the solution
$k_{1,z}=0$ corresponds to the top of the surface mode band, while
above this $k_{x}$ it corresponds to the bottom of the bulk mode band. Similarly, the upper bands depicted in Figs. \ref{fig:sl-p}(c) and Fig. \ref{fig:sl-s}(c) are delimited by the Bragg modes, (\ref{eq:w-g-pi}) and (\ref{eq:w-g-0}).

\begin{figure}
\begin{centering}
\includegraphics[width=8.5cm]{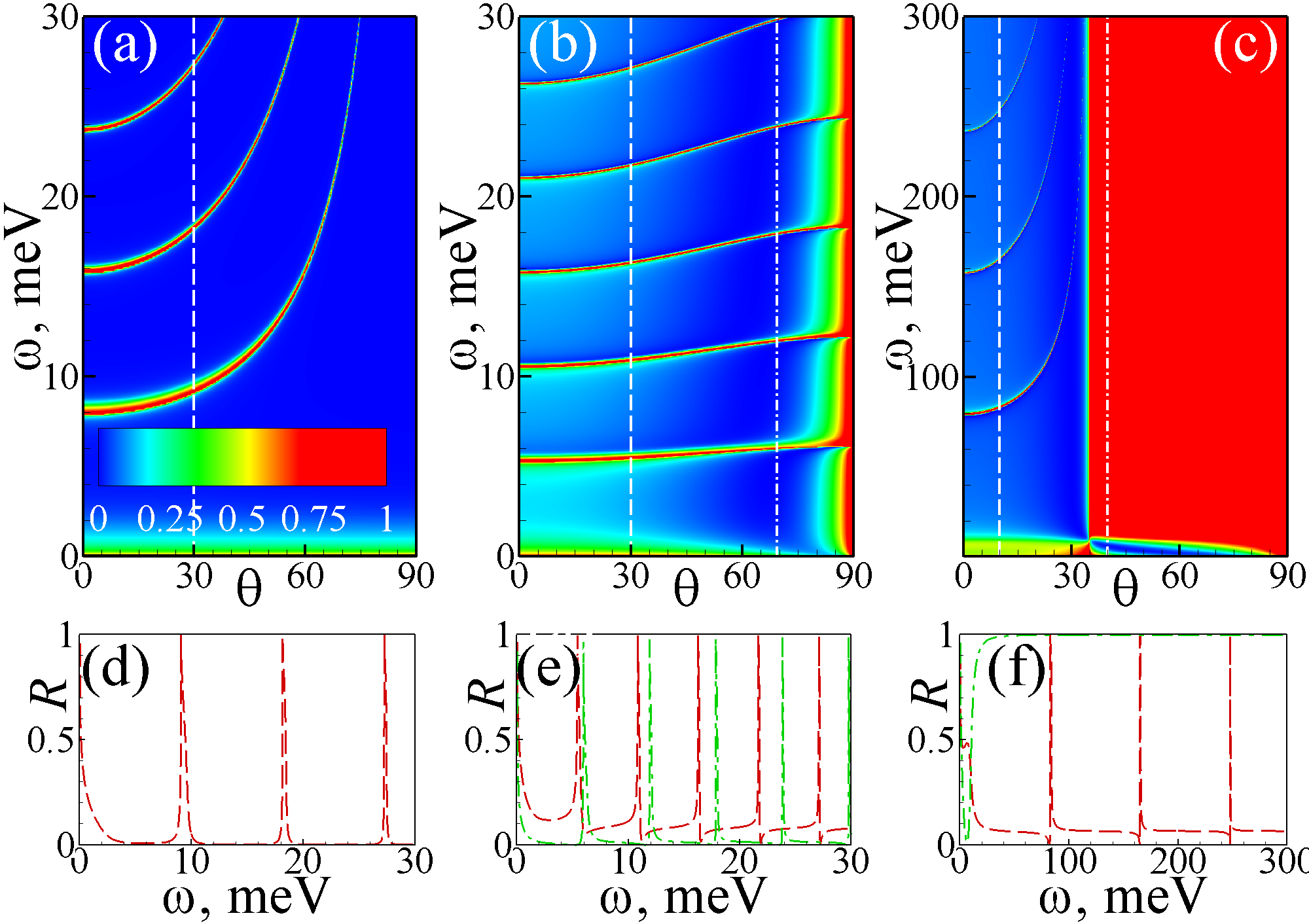}
\par\end{centering}
\caption{$P-$polarization reflectance, $R$, for a semi-infinite graphene
multilayer PC, plotted against the frequency, $\omega$, and the angle of incidence,
$\theta$ (top row), or\textit{ vs} frequency $\omega$ (lower row)
at fixed angles of incidence: $\theta=30^{o}$ {[}dashed line in panel
(d){]}, $\theta=30^{o}$ {[}dashed line in panel (e){]}, $\theta=69.324^{o}$
{[}dash-dotted line in panel (e){]}, $\theta=10^{o}$ {[}dashed line
in panel (f){]}, $\theta=40^{o}$ {[}dash-dotted line in panel (f){]}.
Other parameters are: $\varepsilon_{1}=3.9$, $\Gamma=2.6\,$meV,
$\varepsilon_{2}=3.9$, $E_{F}=0.157\,$eV, $d=40\,\mu\mathrm{m}$
(left column), $\varepsilon_{2}=1.0$, $E_{F}=0.1\,$eV, $d=60\,\mu\mathrm{m}$
(middle column), or $\varepsilon_{2}=11.9$, $E_{F}=0.25\,$eV, $d=4\,\mu\mathrm{m}$
(right column). Notice that the plots (d), (e), and (f) represent the variations along the corresponding vertical lines in panels (a),
(b), and (c), respectively.
\label{fig:reflec-p}}
\end{figure}

At the same time,
changing the graphene Fermi energy, $E_{F}$, e.g. by varying an external gate voltage, it is possible
to tune the width of the gaps, as it can be seen from Figs. \ref{fig:var-gap-eigen}(a) and \ref{fig:var-gap-eigen}(b) (for $p$ and $s$ polarizations, respectively). In particular, the gaps vanish when the Fermi level coincides with the Dirac point.
At the same time, the waveguide modes defined by Eqs. (\ref{eq:w-g-pi}) and (\ref{eq:w-g-0}) remain unchanged because of their above-mentioned independence upon the graphene conductivity.

In order to obtain the expression for the reflectance of an EM
wave from the graphene multilayer stack, we notice that, by virtue
of Eqs. (\ref{eq:disp-p-mat}) and (\ref{eq:disp-s}), the amplitudes $H_{\pm}^{(m)}$
and $E_{\pm}^{(m)}$ are related by:
\[
H_{-}^{(m)}=\rho_{p}H_{+}^{(m)},\qquad E_{-}^{(m)}=-\rho_{s}E_{+}^{(m)}\:,
\]
where
\begin{eqnarray}
\rho_{p}=\frac{\exp\left(ik_{1,z}d\right)\frac{2\pi k_{1,z}}{\omega\varepsilon_{1}}\sigma_{g}}{\exp\left(-ik_{1,z}d\right)\left[1+\frac{2\pi k_{1,z}}{\omega\varepsilon_{1}}\sigma_{g}\right]-\exp\left(iqd\right)},\label{eq:rho-p}\\
\rho_{s}=\frac{\exp\left(ik_{1,z}d\right)\frac{2\pi\omega}{c{}^{2}k_{1,z}}\sigma_{g}}{\exp\left(-ik_{1,z}d\right)\left[1+\frac{2\pi\omega}{c{}^{2}k_{1,z}}\sigma_{g}\right]-\exp\left(iqd\right)},\label{eq:rho-s}
\end{eqnarray}
and Bloch wavevector $q$ for $\rho_{p}$ and $\rho_{s}$ is obtained
from Eqs.(\ref{eq:disp-p}) and (\ref{eq:disp-s}), respectively.
It should be pointed out that, if the graphene conductivity is complex, so is the Bloch wavevector in Eqs. (\ref{eq:rho-p})
and (\ref{eq:rho-s}). Then, applying the above-mentioned
boundary conditions at $z=0$, one can obtain expressions for the
amplitude of the reflected wave in the form:
\begin{eqnarray}
H_{r}^{p}=\frac{\varepsilon_{1}k_{2,z}\frac{1+\rho_{p}}{1-\rho_{p}}-\varepsilon_{2}k_{1,z}+\frac{4\pi}{\omega}\sigma_{g}k_{2,z}k_{1,z}}{\varepsilon_{1}k_{2,z}\frac{1+\rho_{p}}{1-\rho_{p}}+\varepsilon_{2}k_{1,z}+\frac{4\pi}{\omega}\sigma_{g}k_{2,z}k_{1,z}}H_{i}^{p},\label{eq:hr-sl}\\
E_{r}^{s}=-\frac{k_{1,z}\frac{1+\rho_{s}}{1-\rho_{s}}-k_{2,z}+\frac{4\pi\omega}{c^{2}}\sigma_{g}}{k_{1,z}\frac{1+\rho_{s}}{1-\rho_{s}}+k_{2,z}+\frac{4\pi\omega}{c^{2}}\sigma_{g}}E_{i}^{s}.\label{eq:er-sl}
\end{eqnarray}
Notice that, when $\rho_{p}=0$, Eq. (\ref{eq:hr-sl}) coincides with Eq. (\ref{eq:hr})
for the single graphene layer structure. Similarly, when $\rho_{s}=0$, Eq. (\ref{eq:er-sl})
turns into Eq. (\ref{eq:er}).

\begin{figure}
\begin{centering}
\includegraphics[width=8.5cm]{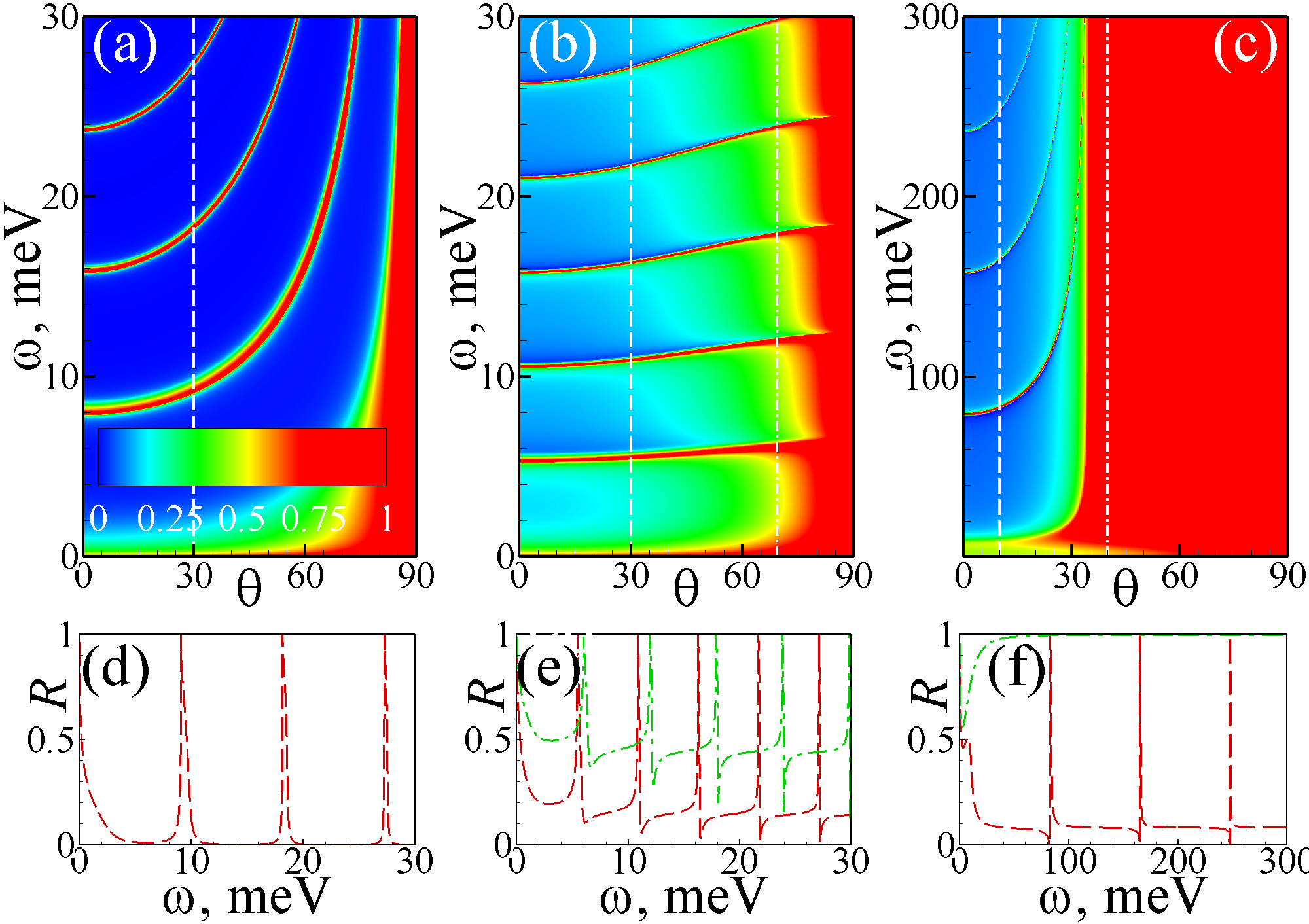}
\par\end{centering}
\caption{Same as in Fig. \ref{fig:reflec-p}, but for $s-$polarized wave.
\label{fig:reflec-s}}
\end{figure}

An incident wave with $\omega$ and $k_{x}$ inside one of the allowed bands of the photonic crystal is (partially) transmitted into the structure.
This effect is clearly seen in Figs. \ref{fig:reflec-p} and \ref{fig:reflec-s} for $p-$ and $s-$polarized waves, respectively).
Thus, when $\omega$ and $k_{x}$ of the incident wave match one of the bands, the reflectance of the graphene multilayer photonic crystal
resembles that of the single-layer graphene {[}compare Figs. \ref{fig:reflec-p}(a), \ref{fig:reflec-s}(a)
with \ref{fig:transmittance}(b), as well as Figs. \ref{fig:reflec-p}(b), \ref{fig:reflec-s}(b)
with \ref{fig:transmittance}(d) and Figs. \ref{fig:reflec-p}(c), \ref{fig:reflec-s}(c)
with \ref{fig:transmittance}(f){]}.

On the contrary, incident EM waves with $\omega$ and $k_x$ belonging to the gaps of the PC band structure induce evanescent waves
(characterized by imaginary Bloch wavevector $q$, in contrast with the PC surface mode with real $q$ and imaginary $k_{1,z}$), and are nearly totally reflected from it.
The graphene-multilayer PC reflectance is considerably higher than that of single-layer graphene heterostructure, and
at certain frequencies can achieve unity {[}see panels (d),(e) and
(f) in Figs. \ref{fig:reflec-p} and \ref{fig:reflec-s}{]}.

\begin{figure}
\begin{centering}
\includegraphics[width=7.5cm]{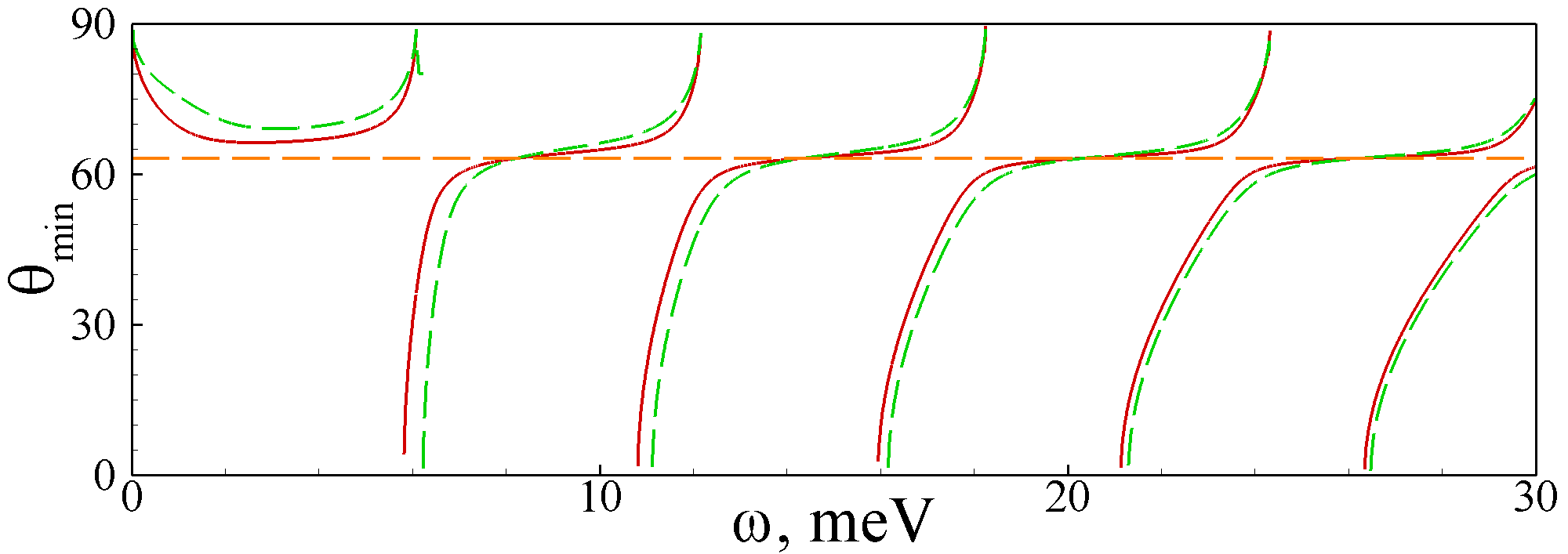}
\par\end{centering}
\caption{Frequency dependence of the angle of incidence corresponding to the
minimal reflectance of $p-$polarized waves, $\theta_{min}$, for two values
of the Fermi level, $E_{F}=0.1\,$eV (solid
lines) and $E_{F}=0.2\,$eV (dashed lines).
Other parameters are the same as for the middle column of Fig. \ref{fig:reflec-p} ($\varepsilon_{2}<\varepsilon_{1}$).
The dashed horizontal line depicts the conventional Brewster angle.
\label{fig:min_angle_sl}}
\end{figure}

Perhaps the most interesting effects take place when $\varepsilon_{2}<\varepsilon_{1}$
{[}panels (b) and (e) in Figs. \ref{fig:reflec-p} and \ref{fig:reflec-s}{]}. As expected, in the vicinity of the Brewster angle of the interface without graphene
($\theta _{br}^0 \approx 63.124^{o}$), the $s-$polarization reflectance exceeds
significantly that of $p-$ polarized waves for all frequencies inside the band {[}compare
dash-dotted lines in Figs. \ref{fig:reflec-p}(e) and \ref{fig:reflec-s}(e){]}, similar to the case of single graphene layer.
However, it is not so for $\omega$ and $k_x$ inside the gaps. Here both
polarizations exhibit an enhanced reflectance. Furthermore, we find some features specific for TM waves.
As it has been shown in the previous section, the presence of graphene at the
interface modifies the angle at which the reflectivity minimum in $p-$polarization occurs and this quasi-Brewster angle ($\theta _{br}$) is frequency-dependent {[}see Fig. \ref{fig:absorbance}(d){]}.
What happens to the minimum reflectivity angle, $\theta_{min}$, when the wave is reflected from the graphene multilayer PC instead of the single interface?
The answer follows from Fig. \ref{fig:min_angle_sl}. When $\omega$ and $k_{x}$ belong to a band of allowed modes, $\theta_{min}$ oscillates around the conventional Brewster angle ($\theta _{br}^0$, dashed horizontal line in the plot), except for the very low-frequency range ($\hbar \omega <3$~meV), where the frequency dependence of the difference, $\theta _{min} -\theta _{br}^0$, resembles that for the single graphene layer structure {[}compare to Fig. \ref{fig:absorbance}(d){]}.
The most striking feature in Fig. \ref{fig:min_angle_sl} is the divergence of $\theta_{min}$ for the frequencies corresponding to the stop--bands of the photonic crystal (compare to the middle column of Fig. \ref{fig:reflec-p}).

The particularity of the situation $\varepsilon_{2}>\varepsilon_{1}$ {[}see panels (c) and (f) in Figs. \ref{fig:reflec-p}
and \ref{fig:reflec-s}{]} is the possibility to excite the $p-$polarized surface mode.
If the angle of incidence is below the critical one ($\theta _c \approx 35^{o}$), the excitation of bulk PC modes takes place, while for $\theta > \theta _c$, only the surface mode can be excited, as it can be seen by the low-frequency minima in the reflectivity spectra {[}see Figs. \ref{fig:reflec-p}(c) and \ref{fig:reflec-p}(f){]}.
A similar spectral shape has been observed experimentally in Ref. \cite{Sreekanth2012}. One can say that the interface between the PC
and the capping dielectric acts as an attenuated total internal reflection
structure for single graphene layer, described in Ref. \cite{BluVasPer2010}.
It should be emphasized that the origin of the low-frequency minimum observed for $s-$polarized waves {[}Figs. \ref{fig:reflec-s}(c) and \ref{fig:reflec-s}(f){]} is completely different. The former is a photonic crystal effect, while the latter  exists also in the single-layer case {[}see Fig. \ref{fig:transmittance}(f){]} and is unrelated to any PC surface mode.

\begin{figure}
\begin{centering}
\includegraphics[width=8cm]{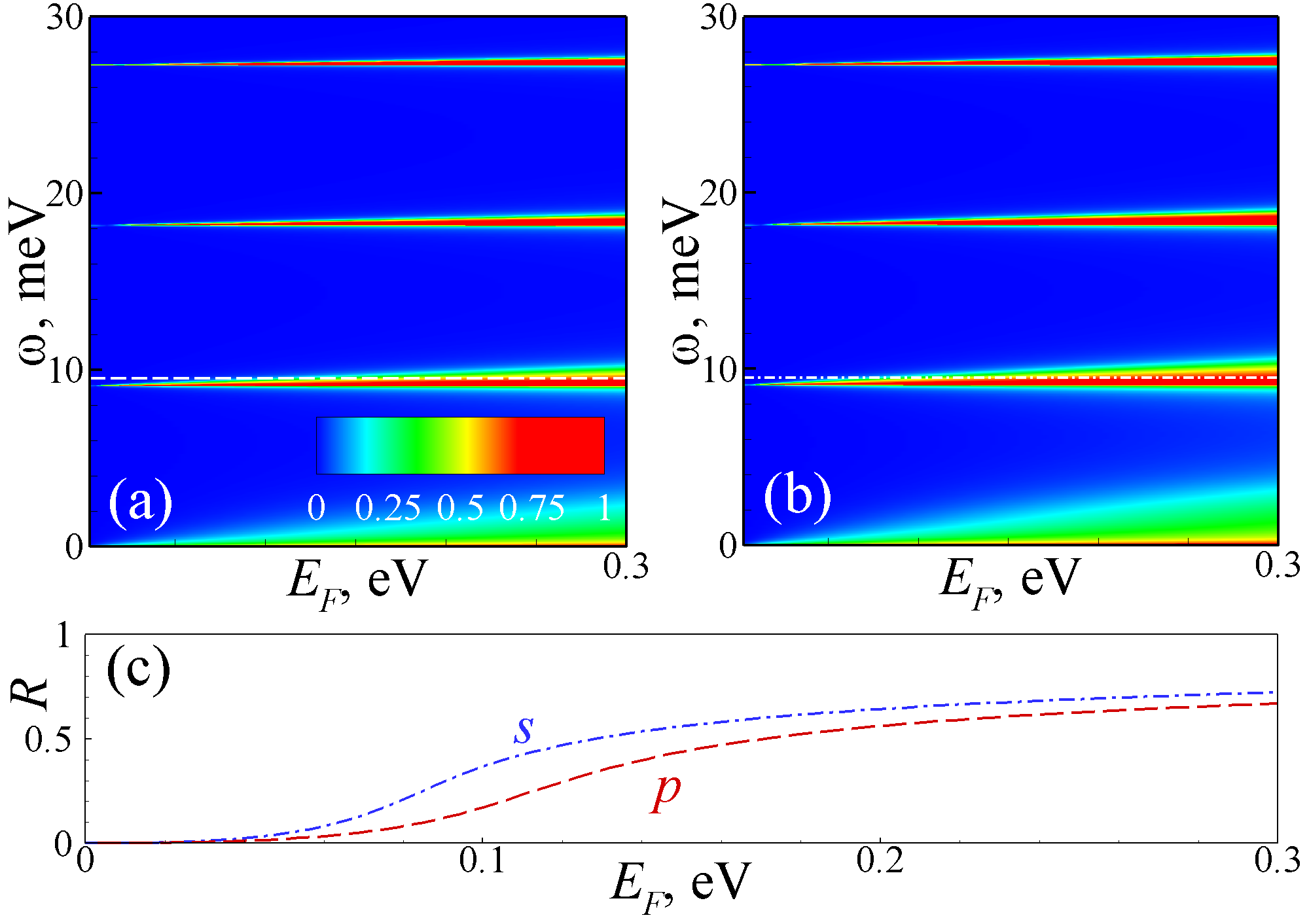}
\par\end{centering}
\caption{(a,b) Reflectance \textit{versus} $E_{F}$ and frequency
for $p-$ (a) and $s-$polarized (b) waves, for the angle if incidence $\theta=30^{o}$.
(c) Reflectance \textit{versus} Fermi level for $p-$ and $s-$polarized waves, for $\hbar \omega=9.5\,$meV [subtracted from panels (a) and (b) along  respective horizontal lines]. Other
parameters are the same as for the left column of Fig. \ref{fig:reflec-p}.
\label{fig:var-gap-sl}}
\end{figure}

The possibility to change gap widths in the graphene multilayer PC
spectrum by changing the Fermi level of graphene layers {[}see Fig.\ref{fig:var-gap-eigen}{]} has an important consequence, the reflectance
of the PC can be dynamically varied through the electrostatic gating,
by changing the voltage applied to the graphene layers. This effect
is depicted in Figs. \ref{fig:var-gap-sl}(a) and \ref{fig:var-gap-sl}(b)
for $p-$ and $s-$polarized waves, respectively. It can be used to design a tunable mirror. One has to choose
the frequency of the incident wave inside one of the allowed bands, for a low
Fermi energy, and inside the gap for a large $E_F$. Then the reflectance
of the structure can be varied in a broad range, as shown in Fig. \ref{fig:var-gap-sl}(c). The dependence
$R\left(E_{F}\right)$ can be made even more abrupt using graphene layers with a smaller damping parameter $\Gamma$.

\begin{figure}
\begin{centering}
\includegraphics[width=8cm]{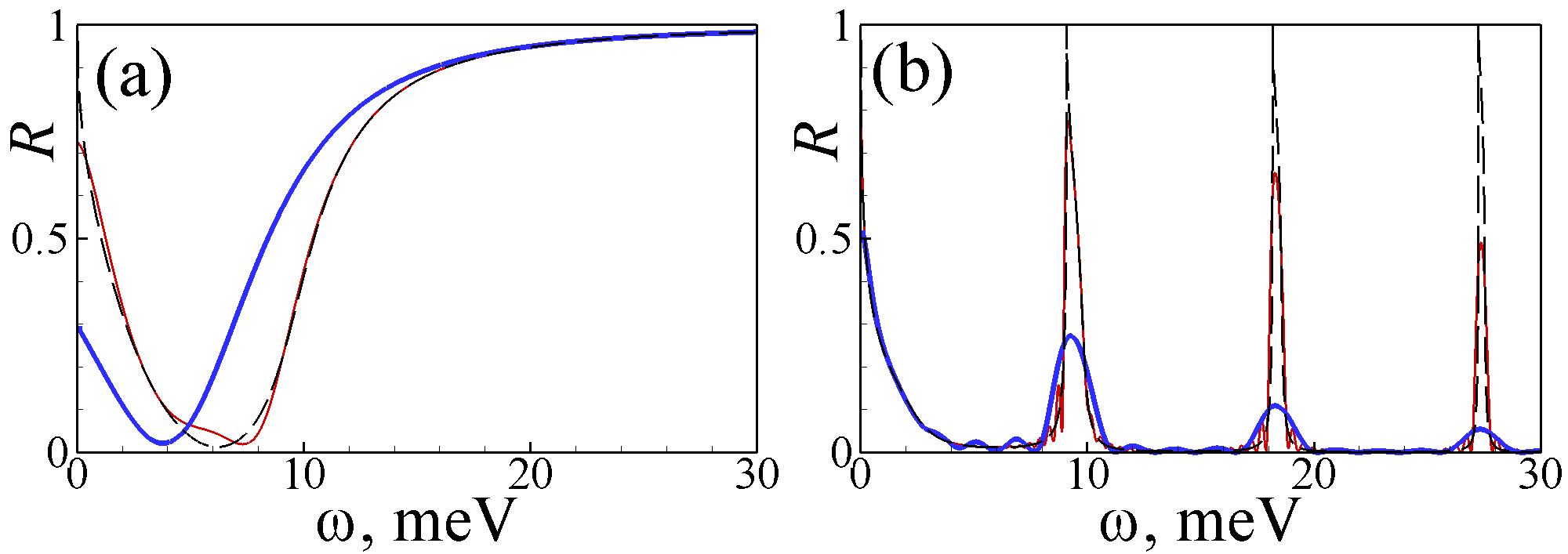}
\par\end{centering}
\caption{Reflectance \textit{vs} frequency for $p-$ (a) and $s-$polarized (b) waves falling on a finite PC containing 5 (thick blue lines) or 20 (thin red lines) graphene layers.
Other parameters are: $\varepsilon_{1}=3.9$, $\Gamma=2.6\,$meV,
$\varepsilon_{2}=3.9$, $E_{F}=0.157\,$eV, $d=40\,\mu\mathrm{m}$, $\theta=30^{o}$ [panel (a)], or
$\varepsilon_{2}=11.9$, $E_{F}=0.25\,$eV, $d=4\,\mu\mathrm{m}$, $\theta=40^{o}$ [panel (b)]. In both panels dashed lines correspond to the case of infinite number of graphene layers in PC for the same parameters.
\label{fig:reflec_fin_lay}}
\end{figure}

All the above results have been obtained for an \textit{infinite} periodic stack of graphene layers. In reality, of course, PCs consist of a \textit{finite} number ($N$) of layers. How does the value of $N$ affect the mode eigenfrequencies and the frequency dependence of the reflectance? As known from the band theory of crystalline solids, the eigenmode spectrum is quantized and corresponds to a discrete set of "allowed" Bloch wavevectors, $q_m=(\pi/d) (m/(N+1));\:m=[1,N]$ obtained from the usual Born--von Karman boundary conditions. For $N\rightarrow \infty $, the Bloch wavevector varies in a quasicontinuous way within the interval $q\in[0,\pi/d]$ and Eqs. (\ref{eq:disp-p}) and (\ref{eq:disp-s}) hold with a very high precision. For relatively small values of $N$, say, $N\sim 10$, the eigenmode band structure is washed away although the density of states retains a qualitative similarity with the case of $N\rightarrow \infty $. The "stop-bands" are broadened and correspond to the maximum reflectivity well below the unity [see Fig. \ref{fig:reflec_fin_lay}(b)], however, the latter increases rapidly with the number of layers, as known for periodically stratified media \cite{Born-Wolf}. Already for $N=20$, the reflectance for $s-$polarized waves is very similar to that for infinite PC [compare dashed and thin solid lines in Fig. \ref{fig:reflec_fin_lay}(b)].

Fig. \ref{fig:reflec_fin_lay}(a) shows the finite size effect on the frequency dependence of reflectance related to the surface mode for $p$ polarization. No qualitative difference between the cases of $N=5$ and $N\rightarrow \infty $ is seen [compare thick solid and dashed lines in Fig. \ref{fig:reflec_fin_lay}(a)], which can be understood by the low dispersion of the surface-type PC mode with respect to the Bloch wavevector [see Fig. \ref{fig:sl-p}(b)]. This mode is, in fact, a Bloch-type surface plasmon-polariton (SPP) excitation induced by the incident wave when the attenuated total reflection conditions are met \cite{PRIMER2013}.
The flatness of the $\omega (q)$ dependence for this PC surface mode originates from the small overlap of the amplitudes of the SPP excitations in the different graphene layers.

\section{Conclusions}

In conclusion, there are several interesting effects related to the optical properties of graphene, which are revealed at oblique incidence.
Some of them are expected already for a single graphene layer or just few of them. Under total internal reflection conditions at an interface between two dielectrics, the presence of graphene leads to EM energy absorption only for $s-$polarized waves. The absorbance attains its maximum exactly at the critical angle of incidence for $s$ polarization (and the maximum value is higher when the graphene conductivity is large), while it vanishes for $p$ polarization [Fig. \ref{fig:absorbance}(c)]. The minimum reflectance of $p-$polarized waves occurs at a (frequency dependent) quasi-Brewster angle that can differ by several degrees from the conventional Brewster angle for the same pair of dielectrics. Close to grazing incidence, graphene (when dielectric constants of substrate and capping layer are equal) is fully transparent to $p-$polarized waves and behaves like a mirror for $s$ polarization.
This effect can be used for polarization-selective guidance of EM radiation.
We have shown that a periodic stack of equally spaced parallel layers of graphene has the properties of a 1D photonic crystal, with narrow stop--bands that are nearly periodic in frequency.
The PC properties are revealed also at oblique incidence. In particular, the stop--bands correspond to singularities of the minimum $p-$polarized reflection angle calculated as the function of frequency, which is an effect of potential interest for optical switching. We investigated the finite PC size effect and found that about 20 periods are sufficient to get the properties very close to those of the infinite PC. Finally, we should stress the possibility of tuning of the gaps' (stop--bands') width by changing the graphene conductivity via electrostatic gating, that would allow for dynamical variation of the reflectance at specific selected frequencies.

\section*{Acknowledgements}

This work was partially supported by FEDER through the COMPTETE Program and by
the Portuguese Foundation for Science and Technology (FCT) through Strategic Project PEst-C/FIS/UI0607/2011.

\bibliographystyle{apsrev}
\bibliography{arbit_angle}

\end{document}